\documentclass[runningheads]{llncs}
\usepackage[T1]{fontenc}
\usepackage{graphicx}
\usepackage[hyphens]{url}
\usepackage{hyperref}
\usepackage{xcolor}
\usepackage{todonotes}

\begin{document}

\title{Autonomic Microservice Management via Agentic AI and MAPE-K Integration}
\titlerunning{Autonomic Microservice Management via AAI and MAPE-K Integration}

\author{Matteo Esposito\orcidID{0000-0002-8451-3668} \and
Alexander Bakhtin\orcidID{0000-0003-3513-7253} \and
Noman Ahmad\orcidID{0009-0005-4228-2493} \and
Mikel Robredo\orcidID{0009-0001-9870-1504} \and
Ruoyu Su\orcidID{0009-0008-6206-8787} \and
Valentina Lenarduzzi\orcidID{0000-0003-0511-5133} \and
Davide Taibi\orcidID{0000-0002-3210-3990}}

\authorrunning{M. Esposito et al.}

\institute{
University of Oulu, Finland \\
\email{\{matteo.esposito, alexander.bakhtin, noman.ahmad, mikel.robredomanero, ruoyu.su, valentina.lenarduzzi, davide.taibi\}@oulu.fi}}

\maketitle

\begin{abstract}
While microservices are revolutionizing cloud computing by offering unparalleled scalability and independent deployment, their decentralized nature poses significant security and management challenges that can threaten system stability. We propose a framework based on MAPE-K, which leverages agentic AI, for autonomous anomaly detection and remediation to address the daunting task of highly distributed system management. 
Our framework offers practical, industry-ready solutions for maintaining robust and secure microservices. Practitioners and researchers can customize the framework to enhance system stability, reduce downtime, and monitor broader system quality attributes such as system performance level, resilience, security, and anomaly management, among others.

\end{abstract}

\keywords{Agentic AI, Large Language Model, Autonomous, Anomalies, Microservices, MAPE, Remediation, Human-Machine Teaming}

\section{Introduction}
\label{sec:intro}
Microservice architecture is an architectural style that structures an application as a collection of small, autonomous services modelled around a business domain \cite{taibilenarduzzi2018}.
Each microservice is designed to perform a specific function and can be developed, deployed, and scaled independently of other services. This approach contrasts with the traditional monolithic architectures, where the entire application is built as a single, interconnected unit.
Nonetheless, according to \cite{raj2023assessing}, ``microservices'' have become an IT buzzword for large enterprise firms. Therefore, tech giants such as Amazon, Azure, and Google promote various services to enable practitioners and researchers to develop, efficiently deploy, and test microservices \cite{sampaio2017supporting}.
However, microservices are likely to have anomalies \cite{nist800-190}. The current state-of-the-art solutions focus on anomaly detection of a distributed architecture by combining trace, logs, and metrics, and performing graph-based deep learning for root cause analysis \cite{zhang2022deeptralog,RCAevalpaper,yu2023nezha,esposito_generative_2025}.

Nonetheless, practitioners still need to spend a lot of manual effort. Moreover, specialized software can suddenly fail, sometimes with far-reaching and chaotic consequences, such as in the recent event involving crowd strike outage\footnote{\url{https://blogs.microsoft.com/blog/2024/07/20/helping-our-customers-through-the-crowdstrike-outage/}}. Amidst state-of-the-art, no autonomous solutions exhibit actual agentic behavior \cite{brumani2022microservices} that can help practitioners to alleviate from their shoulders the weight of managing such complex systems. 

Our approach significantly enhances the resilience and adaptability of microservices environments, addressing faults, emerging threats, and operational challenges. Therefore, our framework can be used to monitor the broader system quality attributes, e.g. performance, resilience, security and anomaly management. Practitioners and researchers can benefit from the following key contributions: (1) We present the first framework that integrates MAPE-K and agentic AI (AAI) concepts, aiming for a robust solution for microservices anomaly management. (2) Introducing the novel concept of an "autonomic threshold", our approach ensures human oversight of high-risk actions in AI-managed systems, balancing autonomy with control~\cite{wright2021measuring,antsaklis2020autonomy}.

\textbf{Paper Structure}. In Section~\ref{sec:background}, we provide a theoretical background for the study. In Section~\ref{sec:motivation}, we provide motivations for our framework, and in Section~\ref{sec:design}, we present its design. In  Section~\ref{sec:limitations}, we discussed the limitations, and in  Section~\ref{sec:ethical}, the ethical implications. In Section~\ref{sec:conlcusions}, we conclude.
\section{Background}
\label{sec:background}
\textbf{An AI Agent}  is a system that can make decisions independently and decide a course of action. In many cases, it emulates human agency \cite{shavit2023practices}. An Agentic AI system can sense its environment, reason through different courses of action, and make decisions to reach specific goals. Furthermore, it should adapt to new information, learn from experience, and respond to changing conditions without the need for human intervention \cite{shavit2023practices}. AAI is proactive; in other words, it initiates actions and does not merely react to explicit commands or input. This is a highly desirable property in many applications where the environment is complex and dynamic, and real-time decision-making and adaptation to constantly changing conditions are required \cite{shavit2023practices}.

Recent studies have addressed the daunting question of liability involving AAIs \cite{10.1145/3593013.3594033}, or instead, its sense of agency. More specifically, \cite{10.1145/3593013.3594033} acknowledges that the agency of algorithmic systems does not alleviate or shift human responsibility for potential algorithmic harm.
 
In the same vein, \cite{LEGASPI2024111298} highlights that understanding how humans' sense of agency (SoA) might be influenced by perceiving control by an AI and thus may be very useful for AI development in general and the spread of human attitudes more positively toward AI, in particular. More precisely, they highlight the research gap related to AI that adapts to changing human SoA in dynamic settings due to the difficulty in modelling and responding to human SoA in intricate settings.

\textbf{MAPE-K cycle} is a framework in autonomic computing for self-managing systems \cite{weyns2013patterns}. Its acronym stands for the following words: Monitor, Analyze, Plan, Execute, and Knowledge. The two major subsystems involved are the manag\emph{ed} and manag\emph{ing} systems. In this context, the AAI implements the framework as the managing system. The second, managed system, is an analyzed system composed of microservices. These two systems interact through sensors and actuators. Sensors collect relevant data that serves as input for the monitoring step. On the other hand, actuators modify the managed system based on the instructions from the execution step. The cycle itself includes the following steps:

\begin{enumerate}
    \item \textbf{Monitor}: Collect system and environment data.
    \item \textbf{Analyze}: Process and interpret the monitored data to identify patterns, trends, and anomalies.
    \item \textbf{Plan}: Develop strategies or actions to address the issues identified in the analysis phase.
    \item \textbf{Execute}: Implement the planned actions to manage the system's behaviour.
    \item[($*$)] \textbf{Knowledge}: Maintain a repository of data, models, and policies that support the other four activities. Also known as the Knowledge Base (KB).
\end{enumerate}

Donakanti et al. \cite{donakanti2024reimagining} previously demonstrated the feasibility of incorporating LLMs into the MAPE-like cycle. However, their approach has several limitations: the authors merged Analyze and Plan stages into a \emph{Synthesize} stage, thus blurring the boundary between the results of the analysis and planning performed based on these results; they allow the framework to execute actions without human oversight; and lastly, they use a monolithic benchmarking system as the evaluation study, while we envision a more distributed and diverse microservice system as the managed system.

\textbf{Layers of microservice-based
systems}. Anomalies in microservices can occur on the different layers of microservice-based systems, each requiring adequate reconstruction \cite{cerny2022microservice}. There are typically three different layers where the potential in-time anomaly detection and remediation actions are required. 

\textbf{1. Static Layer}. This layer concerns the static components of a microservice, e.g., source code. Static analysis is the analysis of the code base of any particular software project to formally verify the target system's correctness, especially when a set of technical reasoning practices is required \cite{cerny2022microservice}. It has long been adopted to support a high-level understanding of legacy monolithic systems in terms of their maintenance and replacement \cite{papotti2012reducing}. For microservice-based systems, static analysis can also help the effectiveness of anti-pattern and bad smell detection \cite{cerny2023catalog}. In addition, the visualization of the microservice structure based on static analysis can also help in terms of anti-pattern identification \cite{cerny2022microvision}. On the other hand, many studies also contributed to the root cause identification of anomalies in microservice architecture \cite{ma2021servicerank}. Moreover, continuous delivery and explainability are considered the main challenges of anomaly detection and failure root causes analysis for microservice \cite{soldani2022anomaly}, which can gain support from Agentic AI. 

\textbf{2. Dynamic Layer}. Furthermore, dynamic analysis can help reconstruct microservice-based systems on different layers. For example, using the telemetry data for system dynamic analysis, it is possible to detect the service dependencies and identify potential architecture smells \cite{cerny2022microservice,al2022using}. Furthermore, the combination of static and dynamic analysis can facilitate the decomposition of monolithic systems into microservices and detect the anomalies therein \cite{krause2020microservice}. The challenges of adopting dynamic analysis for microservice anomaly detection include the compliance and integration of business logic and the interweaving anti-patterns \cite{al2022using}. Though machine learning techniques have been applied for such a purpose \cite{li2021microservice}, Agentic AI can facilitate the practice of improving accuracy and reducing human labour involvement.

\textbf{3. Organizational Layer}. Moreover, the human aspect, i.e., the organizational structure of microservice projects, is another critical layer where anomalies can occur, especially since, according to ``Conway's law'' \cite{conway1968committees}, the system architecture can shift, mirroring the project teams' structure \cite{li2023analyzing}. Therefore, the organizational coupling between microservices and between the corresponding teams can be considered as typical anomalies on the organizational layer since the ideal ``one microservice per team'' status is hard to achieve \cite{amoroso2023one}. To such an end, various context information can facilitate the analysis of developer collaboration and organizational structure optimization \cite{li2024toward,bakhtin2024temporal}. However, this aspect is still limitedly explored, while agentic AI can help evaluate the organizational coupling, identify and recommend the optimal way to decouple, and proactively suggest high-performing collaboration relationships. 
Therefore, we can leverage AAI to facilitate anomaly detection practice in different layers with a specialized MAPE-K loop adopted on each. Introducing new layers, e.g., the energy consumption layer \cite{araujo2024energy}, is also possible in the existing framework by adding a corresponding new loop that targets such anomalies. Previous empirical work has demonstrated the potential benefits on the use of MAPE-K in combination with AI, with techniques such as Markov decision processes~\cite{magableh2020self}, artificial neural networks~\cite{nguyen2017monad} and Long-Short Term Memory (LSTM) models~\cite{de2020data}. However, research on the use of AI in microservice-based systems remains in early stages~\cite{pimentel2021self}.

Our work extends MAPE-K with AAI and a human-in-the-loop (HITL). Cleland-Huang et al. \cite{10.1145/3524844.3528054} proposed the first attempt at inserting HITL in a MAPE-K to address the  Human Machine Teaming (HMT) challenge. Their approach details which tasks in each MAPE-K step the human agent can cooperate with the machine. Conversely, our approach focuses on HMT specifically in the execution plan. Our choice is aimed at averting critical failures of the system, but keeping the managing system as autonomous as possible in handling the managed system.

\subsection{Ansible}
An appropriate number of independent services in a microservices architecture necessitates a corresponding set of automation tools to manage them effectively~\cite{taibilenarduzzi2018}. Among the most critical automation tasks are \textbf{container orchestration} (e.g., \textit{Kubernetes}\footnote{\url{https://kubernetes.io/}}), \textbf{cluster management} (e.g., \textit{Docker Swarm}\footnote{\url{https://docs.docker.com/engine/swarm/}}), and \textbf{system state management} (e.g., \textit{Ansible}\footnote{\url{https://www.ansible.com/}}).

Ansible, introduced in 2012 as an open-source automation platform, leverages YAML-based playbooks to declaratively define system configurations. It operates in an \textbf{agentless}, \textbf{push-based} model over SSH, simplifying deployment across diverse environments. Ansible includes a rich set of built-in modules for managing containers, networks, and services, and also supports the development of custom modules tailored to specific use cases.

\section{Motivation}
\label{sec:motivation}
We consider the point of view of a practitioner tasked with managing a complex system composed of multiple microservices in a highly distributed environment. Such microservices would produce a high volume of execution logs, performance metrics, and service call traces made across the system. Already overwhelmed by the amount of data, the practitioner needs to cope with unexpected failures of different components that can occur, rendering the system unstable. To avert such disastrous failures, there is a need for complexity to be handled in a stronger and automated way \cite{donakanti2024reimagining}.

Incorporating our new framework into the organization's workflows could ease work and enhance the practitioner's performance. Our idea can provide a stable,  self-adaptive and \textbf{proactive} framework for managing anomalies in microservices.

\textbf{Monitoring.} The AAI continuously monitors the microservices environment, collecting data from execution logs, performance metrics, and service calls. Using advanced AI algorithms, it analyzes this data in real-time to detect anomalies and potential threats before they escalate into major issues, thus anticipating the failure of the system.

\textbf{Proactive Fault Management.} When anomalies are detected, the Agent not only alerts the practitioner but also provides actionable insights and recommended solutions. This proactive fault management \textbf{reduces the time and effort} required to detect and remediate issues by anticipating their occurrence, thus allowing the practitioner to focus on more strategic tasks.

\textbf{Customizable Solution.} Our framework provides a flexible foundation that practitioners can build upon, considering their specific needs. Such adaptability means that the framework can change with the company's needs: whether it is adapting to new services, changing performance standards or new threats, \textbf{all without expensive re-engineering}.

Overall, integrating our framework into the company's practices will \textbf{transition} the role of a practitioner from \textbf{a reactive firefighter} to a \textbf{proactive strategist}. Now, the practitioner will primarily be directed by this framework in daily routine monitoring and anomaly detection, which leads them to pay attention to higher-level planning and optimization with consequent improvements in resilience and efficiency.

\section{Design}
\label{sec:design}
\begin{figure}[t]
    \centering
    \includegraphics[width=\linewidth]{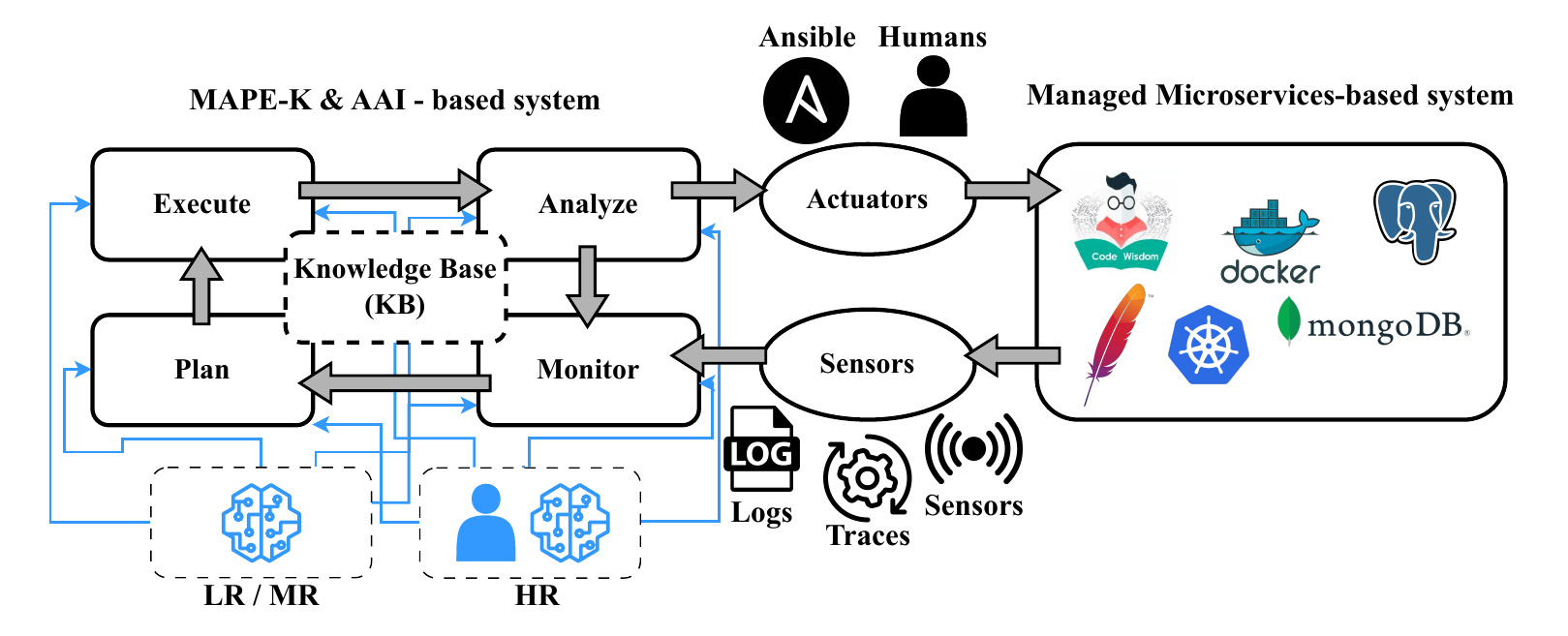}
    \caption{Diagram on the interaction between AAI and MAPE-K components.}
    \label{fig:mape_k}
\end{figure}

This section presents the envisioned framework for autonomous anomaly detection and remediation leveraging Agentic AIs.
We designed our framework around the MAPE-K cycle. We describe for each step of the cycle how we contextualize it to address the challenge of autonomous anomaly detection and remediation. Moreover, for each step, we propose the implementation methodology and the relationship with the knowledge base, and summarize the contribution of AAI. Figure~\ref{fig:mape_k} displays the diagram showing the interaction between AAI and MAPE-K components.

\subsection{Monitor}
\textbf{Context}. The monitoring stage aims to supervise the development and operation activities of the microservice-based systems by continuously collecting data from multiple layers, e.g., source code, system operation, and organizational structure. The whole process should be conducted efficiently when sufficient data is collected. 

\noindent\textbf{Method}. Changes in the source code as measured by the critical metrics can be monitored through the updates in the repositories of version control systems by continuously crawling data via public APIs and using customized tools or via CI/CD actions. Open tracing tools can facilitate logging and tracing of the system operations on the dynamic layer. 

\noindent\textbf{Knowledge}. The knowledge gathered in this stage consists of a large volume of raw data from the different layers of microservice-based systems. Such data shall be further processed and analyzed for input by the AAI in the later stages. Meanwhile, the output of AAI shall also be continuously monitored. 

\noindent\textbf{Agentic AI}. The AAI shall facilitate the monitoring stage by selectively collecting useful data and automatically categorizing the data, targeting potential tasks on different layers. Furthermore, the AAI shall also facilitate the coordination of all integrated tools adopted. Human interaction shall be allowed when new metrics are introduced and corresponding monitoring instructions are deployed.

\subsection{Analyze}
\textbf{Context}. The analysis stage consists of handling the data collected in the monitoring stage, making it processable, producing interpretable outcomes to be added to the KB, and, therefore, considering it in the planning stage to make data-driven decisions. Similarly, the entire process shall maintain the data pipelines 
 of the three layers generated in the previous stage.

\noindent\textbf{Method}. The analysis stage initially pre-processes the raw data collected through the monitoring stage. The framework will consider diverse data pre-processing techniques stored in the KB to make the raw data obtained through the suggested three layers analyzable. For instance, we can detect the impact of the implemented changes in the source code through a wide variety of methods based on anomaly detection, such as change point detection techniques for CI performance control, as well as multiple ML algorithms for defect prediction. Furthermore, ML ensemble learning techniques and deep learning techniques can help in fault prediction and localization tasks involving time-series-based continuous data flow, such as log data generated through log tracing tools. In the context of the organizational layer, automated analysis of collaboration relationships through Social Network Analysis has proven suitable for showing developers' collaboration patterns \cite{li2023analyzing,bakhtin2024temporal}. 

\noindent\textbf{Knowledge}. Within the analysis stage, the KB is crucial to obtaining data-driven results. The KB provides the monitored raw data, which is further preprocessed and analyzed. Similarly, the KB is the source from which the AAI chooses and adopts the most suitable analysis techniques based on their goodness of fit for each layer. Furthermore, the KB is the storage where the AAI stores the detected anomalies and potential future predictions to be interpreted in the planning stage. The Agent's continuous learning mechanism should leverage the knowledge stored in the KB in this stage to provide exhaustive data-driven decisions and robust reinforcement learning of the models used in the KB for future analysis cycles.

\noindent\textbf{Agentic AI}. The AAI deployed in this stage shall provide the most efficient data analysis pipelines existing in the KB, given the nature of the metrics identified in the data categorization of the monitoring stage. The AAI should calculate the best analysis methods to cope with the potential insights extracted from the data, such as anomaly detection or collaboration optimization. Human interaction should be included when AAI presents low performance on the chosen models and detects anomalies in the integrated analysis pipelines' accuracy results. 

\subsection{Plan}
\textbf{Context}. The planning step is the core of the MAPE-K cycle \cite{1160055}.  In our context, we tailored the tasks for which the plan step is responsible as follows: \textit{Goal Setting}: Define the goals and desired state of the managed system, excluding strategy development in this context; \textit{Resource Allocation}: Decide on resource allocation, scheduling, and roles for plan implementation; \textit{Risk Assessment}: Assess potential risks and impacts, and develop mitigation strategies;  \textit{Action Plan Formulation}: Formulate an action plan, often involving human input, as directives for actuators; \textit{Feedback Integration}: Gather feedback from previous cycles or external sources to ensure continuous improvement and adaptation.

\noindent\textbf{Method}. The AAI will collect the results of the analysis step, particularly the identified anomalies, from the knowledge base to formulate the action plan. To allow autonomous planning and execution, we design this step, specifically the AI agent, to formulate the plan in a machine-understandable format. For simplicity, we can refer to technologies such as Ansible Playbooks. In our context, risk assessment is crucial to avoid a catastrophic failure. Previous research showed that AI models such as LLMs can perform preliminary security risk analysis \cite{esposito2024leveraging,esposito_large_2024}; therefore, AAI can leverage LLM to evaluate the risk associated with the formulated plan. Moreover, this evaluation is pivotal for the execution step and for selecting the correct type of actuators.

The risk associated with the threshold does not have a consequence on the overall design of the framework. For presentation purposes, we define the following risk level:   

\textit{Low Risk} (LR): High-level actions like parameter tuning and log cleanups that, if incorrect, don't cause system instability.

\textit{Medium Risk} (MR): Actions affecting system performance, such as increasing virtual memory, reinstating services, and backups. Incorrect actions may cause minor instability but can be recovered.

\textit{High Risk} (HR): Low-level actions like key distribution, certificate management, service reinstatement, and system upgrades. Incorrect actions can cause significant instability.

\begin{figure}[htb]
    \centering
    \includegraphics[width=0.8\linewidth]{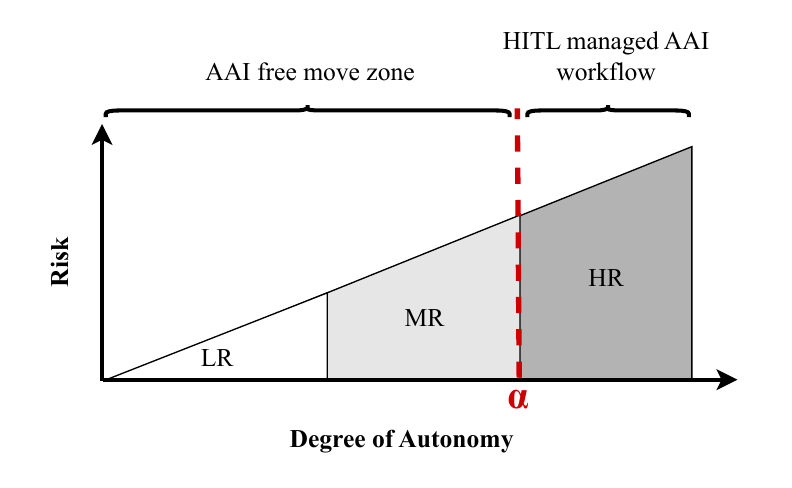}
    \caption{Overall risk associated framework design.}
    \label{fig:autonomic_threshold}
\end{figure}

According to our proposed sample risk levels, an AAI can perform LR and MR actions freely, directly issuing commands through the CLI actuators. Conversely, we should consider a different approach to HR actions. Allowing an AAI to control the managed system without any ``human-in-the-loop'' for such actions is risky. Therefore, we should define an ''autonomic threshold'' ($\alpha$). Previous works have highlighted the importance of categorizing the ''degree of autonomy'' within a software system~\cite{wright2021measuring}, as well as the measurement on the ''level of autonomy'' of an autonomous system~\cite{antsaklis2020autonomy}. We propose a threshold based on how many HR actions are associated with each action formulated in the plan. We can also define subtypes of actions and compute a weighted sum of the HR actions based on these subcategories. When the value $\alpha$ is reached, AAI should stop taking actions and alert a human agent to take those on its behalf (Figure~\ref{fig:autonomic_threshold}).

\noindent\textbf{Knowledge}. In the planning step of MAPE-K, the knowledge base is instrumental in providing the relevant information.
AAI extracts data points regarding goal setting, resource allocation, risk assessment, action plans, and feedback integration from the KB. In particular, the knowledge base contains data from the analysis stage, which is vital to obtaining information regarding the anomalies and ensuring the action plan is oriented toward efficient remediation of the detected anomalies.

\noindent\textbf{Agentic AI}. AAI is vital in considering and testing more than one strategy or action plan. This is done using optimization algorithms and scenario simulations during the planning process. The AAI could rate them according to their potential impact and the likelihood of their success in addressing current and future objectives.

\subsection{Execute}

\textbf{Context}.
The execution stage is where the action plans developed at the planning stage are executed. In this phase, the managed systems are modified to remediate the detected anomalies and bring them to the desired state defined by the KB goals. This phase uses the software actuators to execute the planned actions, and potentially requests inputs and decisions from the human user depending on the value of $\alpha$.

It is worth noticing that, in our context, \textbf{actuators} are of two types: human and software. The framework is designed to be as autonomous as possible, but no current definition of 'autonomic threshold' exists, i.e., how to decide when the HITL is needed. Finally, allowing an AAI to fully control the cloud architecture has non-negligible risks that we must address. 

\noindent\textbf{Method}. The execution step fetches the action plan from the knowledge base. The AAI coordinates the planned actions based on the identified risk levels. The AAI generates the appropriate plan instruction, e.g., the Ansible playbooks, and sends the action commands via the CLI actuators for executing low and medium-risk actions. For high-risk actions, the AAI will alert human operators when it reaches the autonomic threshold, $\alpha$, and requires human intervention to execute such actions.  Moreover, execution logs and outcomes are monitored and recorded in real-time to provide feedback and populate the knowledge base for the next cycle. 
We note that, while we describe Ansible within our framework, any comparable automation tool can be used either as a substitute for or in conjunction with Ansible, depending on the system requirements and context.

\noindent\textbf{Knowledge}. The knowledge base is the source of action plans and historical data on past executions. The ongoing execution stage updates the KB with logs of success and failure, performance metrics, and deviations compared to expected results. This information is critical to shape future MAPE rounds and, therefore, enhance the autonomic capabilities of the system.

\noindent\textbf{Agentic AI}. The AAI must ensure the correct execution of the planned actions, infer the risk level, and support human interaction in the case of HR actions to assess critical decisions before implementation.

\section{Limitations}
\label{sec:limitations}
This section presents our approach's challenges, limitations, and possible solutions.

\subsection{Monitor}  

\noindent
\textbf{Limitation 1M} Data collected from different layers, such as source code, the system operation, and the organizational structure, is very high in volume and diverse in structure; it may easily overwhelm the system and render the data processing inefficient, leading to a bottleneck.  \textbf{Solution:} We will provide data filtering and aggregation algorithms during the data collection and KB storage to reduce the amount of data sent for further processing, and use distributed data processing frameworks to handle big data.

\subsection{Analyze}
\textbf{Limitation 1A} The produced data volumes are big, which may compromise the accuracy of the anomaly detection and fault prediction models, hence having a high false positive or negative rate. \textbf{Solution:} We can use ensemble techniques by combining multiple models to detect anomalies via majority voting. Continually update and retrain such models with new data to enable them to continue learning and responding to the system changes.

\noindent
\textbf{Limitation 2A} The analyzed data is usually of a very high dimension, so it is cumbersome to handle and understand; hence, it is very likely to skew the analysis results. \textbf{Solution:} Dimensionality Reduction Techniques, such as Principal Component Analysis (PCA) or t-distributed Stochastic Neighbour Embedding (t-SNE), can be used to reduce the dimensionality of the data while retaining relevant information.

\subsection{Plan}  
\noindent\textbf{Limitation 1P} It is hard to know which part of the system should be automated and which would still require human intervention, especially in the case of high-risk actions that may affect the reliability and safety of the system. \textbf{Solution:}
We propose the adjustable autonomic thresholds that depend on the system performance and the confidence level of the AAI.

\subsection{Execute}  
\noindent\textbf{Limitation 1E} System errors or conflicts can lead to unpredictable outcomes due to the automation at the 'execute' stage, which can make the system unstable. \textbf{Solution:} The AAI must be able to perform rollbacks to restore the previously known stable point of the system, or request such an action from the human user.

\subsection{General Considerations}

\noindent\textbf{Limitation 1G} Catastrophic AAI actions can occur due to unforeseen events and result in critical system failures. Such is the situation when \textbf{AAI can keep remediating an issue caused by itself}, causing a degrading feedback loop. \textbf{Solution:} Our implementation will include checks for such degradation loops. Moreover, the AAI should create an incident response plan that covers automatic rollback procedures and human intervention protocols for detecting and mitigating such events by restoring the system to its former stable condition.
\noindent\textbf{Limitation 2G} There is no historical data on autonomic AAI-based anomaly detection and remediation systems for microservices. \textbf{Solution:} Our proposed approach faces novel challenges no prior study has faced. Thus, the future empirical validation will lay the cornerstone of AAI for microservice anomaly detection and remediation.
\section{Ethical Implications}
\label{sec:ethical}
 We live in a fast-changing world. Until recently, AI mostly categorized things, but now it enhances human imagination. According to \cite{10372461}, SE researchers and practitioners should not ``look away."  Leveraging AI in the context of software engineering, although not involved with human data directly, still carries ethical risks and implications \cite{10372461}. We identified the following ethical considerations that we should address when implementing and deploying the proposed framework:

\noindent
\textbf{Accountability and Transparency}. With AAI making autonomous decisions, clear accountability for errors or unintended consequences is crucial. We will consider implementing transparency mechanisms like detailed logging and decision tracing to allow human agents to understand AI decisions.

\noindent 
\textbf{Bias and Fairness}. AI decisions are only as fair as the data and algorithms used within. We will design the feedback loop in the KB to periodically audit data and performance to flag and reduce any biases. This issue links back to the scarce data constraint we highlighted earlier. 

\noindent\textbf{Privacy and Data Security}. Continuous data collection and analysis raise privacy and security concerns, especially when considering the organizational layer. We will consider robust data encryption and access control measures and obey data protection regulations to prevent privacy and security breaches on both the managing and managed systems.

\section{Conclusions}
\label{sec:conlcusions}
In this vision paper, we presented our framework, which is the first to unify the autonomic computing theory and the MAPE-K cycle with the novel scenarios enabled by the agentic AI in the context of microservice architecture. Our framework allows practitioners to focus on higher-level microservice system design and management while only being in the loop of monitoring and remediation for making critical decisions which cannot be responsibly allocated to agentic AI. On the other hand, researchers can build on our pioneering study and the results of our future empirical validation to enhance current autonomic systems. Our future research effort will focus on implementing the proposed framework, empirically validating it, and overcoming the highlighted limitations.

\section*{Acknowledgments} 
This work has been funded by the Research Council of Finland (grants n. 359861 and 349488 - MuFAno), by Business Finland (grant 6GSoft\cite{akbar_6gsoft_2024}), and FAST, the Finnish Software Engineering Doctoral Research Network, funded by the Ministry of Education and Culture, Finland.
\bibliographystyle{splncs04}
\bibliography{main}
 
\end{document}